\documentclass[conf]{new-aiaa}
\usepackage[utf8]{inputenc}

\usepackage{graphicx}
\usepackage{amsmath}
\usepackage[version=4]{mhchem}
\usepackage{siunitx}
\usepackage{longtable,tabularx}
\usepackage[hyphens]{url}
\usepackage{multirow}
\usepackage{makecell}
\usepackage{subfig}
\title{Regional Air Mobility Flight Demand Modeling in Tennessee State}

\author{Kamal Acharya\footnote{Graduate Student, Department of Information Systems, 1000 Hilltop Circle, Baltimore, MD 21250, USA.} Mehul Lad\footnote{Graduate Student, Department of Information Systems, 1000 Hilltop Circle, Baltimore, MD 21250, USA.} and Houbing Song\footnote{Associate Professor, Department of Information Systems, 1000 Hilltop Circle, Baltimore, MD 21250, USA.}}
\affil{University of Maryland Baltimore County,
Baltimore, Maryland, 21250, USA}
\author{Liang Sun\footnote{Associate Professor, Department of Mechanical Engineering, One Bear Place \#97356, Waco, TX 76798-7356, USA .}}
\affil{Baylor University, Waco, TX, 76798-7356, USA. AIAA Senior Member. }

\begin{document}

\maketitle

\begin{abstract}
Advanced Air Mobility (AAM), encompassing Urban Air Mobility (UAM) and Regional Air Mobility (RAM), offers innovative solutions to mitigate the issues related to ground transportation like traffic congestion, environmental pollution etc. RAM addresses transportation inefficiencies over medium-distance trips (50-500 miles), which are often underserved by both traditional air and ground transportation systems. This study focuses on RAM in Tennessee, addressing the complexities of demand modeling as a critical aspect of effective RAM implementation. Leveraging datasets from the Bureau of Transportation Statistics (BTS), Internal Revenue Service (IRS), Federal Aviation Administration (FAA), and other sources, we assess trip data across Tennessee's Metropolitan Statistical Areas (MSAs) to develop a predictive framework for RAM demand. Through cost, time, and risk regression, we calculate a Generalized Travel Cost (GTC) that allows for comparative analysis between ground transportation and RAM, identifying factors that influence mode choice. When focusing on only five major airports (BNA, CHA, MEM, TRI, and TYS) as RAM hubs, the results reveal a mixed demand pattern due to varying travel distances to these central locations, which increases back-and-forth travel for some routes. However, by expanding the RAM network to include more regional airports, the GTC for RAM aligns more closely with traditional air travel, providing a smoother and more competitive option against ground transportation, particularly for trips exceeding 300 miles. The analysis shows that RAM demand is likely to be selected when air transportation accounts for more than 80\% of the total GTC, air travel time is more than 1 hour and when the ground GTC exceeds 300 for specific origin-destination pairs. The data and code can be accessed on GitHub. \footnote{Github: \url{https://github.com/lotussavy/AIAAScitecth-2025.git }}

\end{abstract}

\section{Nomenclature}

{\renewcommand\arraystretch{1.0}
\noindent\begin{longtable*}{@{}l @{\quad=\quad} l@{}}
$C_m$& Cost of trip per mile per passenger for \textit{m} mode of transportation \\
$d$& Driving distance derived from the Google Map Distance Matrix API Service \\
$GTC_m$ & Generalized travel cost for \textit{m} mode of transportation \\
$m$ & Mode of transportation(Ground(G) or Airlines(A) or Regional Air Mobility(RAM)\\
$P_m$ & Probability of selecting \textit{m} mode of transportation \\
$R_m$ & Risk of trip for \textit{m} mode of transportation \\
$T_m$ & Time of trip for \textit{m} mode of transportation \\
$U_m$ & Utility for \textit{m} mode of transportation \\
$VSL$ & Value of Statistical Life(VSL)  is the amount society is willing to pay to save a life\\
$W$  & Average median hourly wage of origin and destination Metropolitan Statistical Areas(MSA) \\
$\alpha_m$ & Number of fatalities per mile for \textit{m} mode of transportation \\
$\beta$ & Standard mileage rates provided by the United States Internal Revenue Services \\
\end{longtable*}}

\section{Introduction}
\lettrine{T}{ransportation} inefficiencies, particularly over medium-distance trips ranging from 50 to 500 miles, present a significant challenge to modern mobility systems. Traditional ground transportation modes often face constraints such as congestion and extended travel times\cite{INRIX2024}, while conventional air transportation is limited by the availability and accessibility of major airports, resulting in underserved routes within this distance range. To avoid the challenges of car travel during peak traffic times, individuals often seek alternative transportation methods, such as using city subways or adjusting their work schedules to travel during off-peak hours. Although these alternatives generally involve less stress and shorter travel times, they can be more complex, inconvenient, and sometimes uncomfortable. Advanced Air Mobility (AAM) has emerged as a promising solution to bridge these gaps.

AAM represents an emerging aviation network that employs innovative aircraft and a range of cutting-edge technologies to transport individuals and goods securely, efficiently, economically, and eco-consciously to local destinations, thus connecting communities inadequately served by existing transportation methods\cite{johnson2022nasa}. AAM is divided into Urban Air Mobility (UAM) and Regional Air Mobility (RAM). UAM focuses on transporting people and cargo within cities, while RAM operates on a regional level\cite{antcliff2021regional}. Our study concentrates on RAM, specifically in Tennessee state. RAM aims to utilize electric and autonomous aircraft for urban and regional transportation and seeks to provide flexible air mobility, cargo delivery, and emergency services through a connected multimodal network. With rapid advancements in electric propulsion and autonomous systems, the RAM market emphasizes sustainability and efficiency to address urban congestion and environmental issues. The implementation of RAM services presents multiple challenges and interdisciplinary constraints that companies must address, including:
\begin{enumerate}
\item Decisions regarding the placement of air taxi stations. 
\item Effective routing and coordination of thousands of air taxis across the network. 
\item Advanced data analytics to forecast demand in real-time. 
\item Minimize passenger commute time and costs, maximize the efficiency of on-demand ride-sharing.
\item Operational issues such as developing pricing strategies, evaluating first and last mile delivery options, and monitoring critical metrics such as the status of air taxi batteries and maintenance needs.
\end{enumerate}

The purpose of this study is to provide insights into the demand for RAM flights, focusing on its  feasibility. To achieve this objective, the initial step involves assessing the demand for RAM services by analyzing existing trip demand data and identifying trips that qualify for RAM (ranging from 50 to 500 miles)\cite{mckinseyShorthaulFlying}. The trip demand is examined at the Metropolitan Statistical Areas (MSAs) level within Tennessee. Using various datasets from sources such as the Bureau of Transportation Statistics (BTS), Internal Revenue Service (IRS), and the Federal Aviation Administration (FAA), we conducted cost, time, and risk modeling. Based on these models, we calculated the Generalized Travel Cost (GTC) to predict whether a trip was served by ground transportation or RAM. We selected five regional airports—BNA, CHA, MEM, TRI, and TYS as hubs for RAM operations. We also analyzed the impact of adding other smaller airports in the switching rate of people from ground transportation to the RAM flights.

There is a limited amount of research focusing on demand prediction for RAM. Most existing studies \cite{rajendran2019insights}\cite{rajendran2021predicting}\cite{ahmed2024demand} concentrate on UAM demand prediction and rely solely on a single dataset of yellow taxi trips from New York City. In contrast, research on RAM demand is scarce. Only one notable study\cite{justin2021demand} has modeled RAM demand, predicting demand for the year 2040 by tuning data from the Federal Highway Administration (FHWA)\footnote{FHWA: \url{ https://www.fhwa.dot.gov/policyinformation/analysisframework }.Last accessed on [May 14\textsuperscript{th} 2024]} published in 2008. Existing studies either focus on UAM demand using the New York City dataset or on fixed-year RAM demand predictions. Our research assigned trip-demand generated to RAM demand based on cost, time, and risk factors, which previous studies have overlooked. We also mapped trip generation from each MSA to RAM demand and analyze the impact of introducing new airports as hubs on the shift from ground transportation to RAM travel. This research is the first to use MSA-level demand forecasting and to concentrate on the Tennessee region.

The rest of this paper is organized as follows. In Section III, we summarize the previous related research that has been conducted. Section IV provides the methodology of the research, presenting the technical approach used for modeling the demand for RAM. This section details the cost modeling, time modeling, and risk modeling involved in ground transportation and airlines, along with the datasets used and the region of interest, Tennessee State. Section V presents the results obtained and how demand shifts from ground transportation to airlines. Finally, in Section VI, we conclude with the main contributions of our research.

\section{Literature Review}

There has been a limited focus on RAM demand modeling, with most research predominantly centered on UAM. However, a few studies have explored RAM, leveraging its unique characteristics and challenges. 

The study \cite{rajendran2019insights} proposed a two-phase clustering method for optimizing air taxi station placement, the first phase identifying high-demand areas via multimodal transportation data and the refinement phase addressing demand and feasibility constraints. Expanding on this, study \cite{rajendran2021predicting} integrated environmental variables such as temperature and wind speed to enhance forecasting accuracy. Employing machine learning models, the study classified UAM demand into low, moderate, and high categories, leveraging k-means clustering to optimize resource allocation across New York City. Similarly, research work \cite{ahmed2024demand} applied deep learning techniques, using a "demand ratio" metric derived from New York City taxi data\footnote{Yellow Taxi Trip Data: \url{ https://catalog.data.gov/dataset/2023-yellow-taxi-trip-data }.Last accessed on [August 30\textsuperscript{th} 2024]} to capture temporal and spatial demand patterns. The research work \cite{rimjha2021commuter} used a mixed conditional logit model to estimate UAM demand in Northern California, emphasizing the influence of fare levels and vertiport accessibility. Another research \cite{rimjha2022impact} analyzed the impact of airspace restrictions on UAM in New York City, highlighting the need for policy interventions to optimize vertiport placement. A study in Chengdu, China \cite{qu2024demand} focused on UAM demand integration with urban transit systems, employing a four-step model to forecast demand for 2030.

Recent advancements in electric propulsion and autonomy have emerged as transformative solutions for RAM, offering opportunities to improve operational efficiency and expand connectivity. The study by \cite{roy2021future} emphasized the transformative potential of advancements in electric propulsion and autonomy for RAM, focusing on ride-sharing and real-time data integration to enhance market attractiveness. Their computational framework incorporated curve-fitted models to compare RAM with traditional airlines and automobiles, identifying RAM’s potential for operational efficiency and improved connectivity. Similarly, \cite{bridgelall2023forecasting} adopted hybrid methodologies that combined top-down and bottom-up approaches for demand forecasting across distance bands of 100–400 miles. Using GIS and network trimming techniques, the study analyzed over 2,000 U.S. routes, revealing that shorter routes required higher capital investment but offered scalability for regional implementation.

Addressing the inefficiencies in regional air travel, \cite{villa2023business} proposed cost-efficient aircraft designs with optimized load factors and operating models to improve connectivity, particularly in underserved regions. They highlighted operational inefficiencies in the current state of U.S. regional air travel and identified solutions to enhance its viability. Furthermore, Justin et al. \cite{justin2022integrated} developed a hierarchical multi-objective optimization framework that balanced profitability and emissions for fleet assignments and scheduling in the U.S. Northeast Corridor. Another study by Justin et al. \cite{justin2021demand} integrated demand modeling with a multinomial logit approach, optimizing aircraft assignment and schedules to achieve sustainability and profitability. This work showcased the potential of RAM to leverage battery advancements and adjust networks for sustainable regional connectivity. The study \cite{acharya2024demandmodelingadvancedair} employed a methodology similar to ours; however, the key differences lie in their focus on census tract-level analysis, which examines smaller geographic sections, and their consideration of trips exclusively based on employment.

Despite these contributions, the majority of existing research does not adequately incorporate risk factors into RAM demand modeling or address the complexities of large regional areas such as MSAs. These gaps highlight the need for further exploration of RAM-specific demand modeling frameworks to better capture the nuances of regional connectivity and address the broader challenges unique to RAM.
\section{Methodology}
\begin{figure}[hbt!]
\centering
\includegraphics[width=0.6\textwidth]{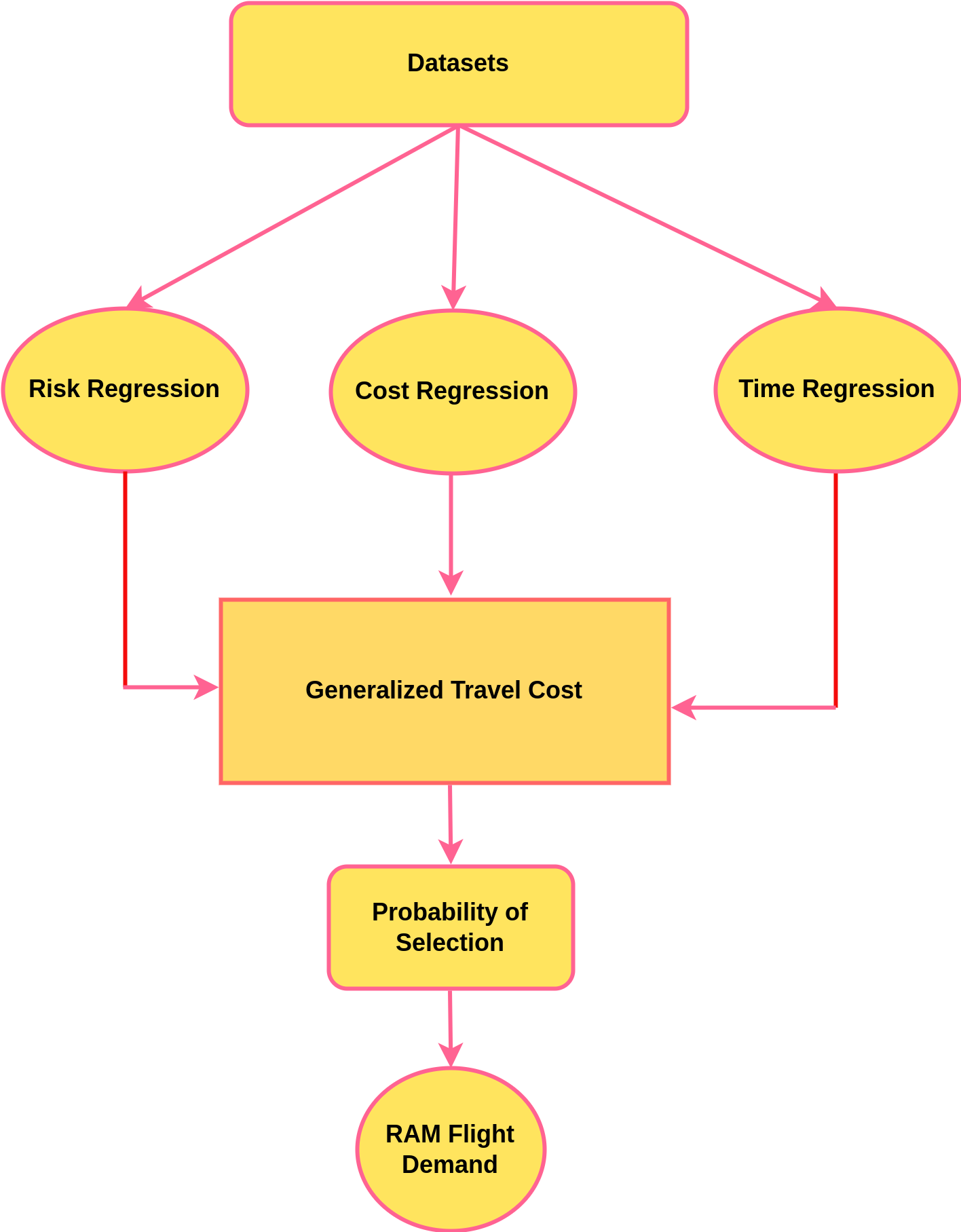}
\caption{Approach for RAM demand modeling}
\label{fig:Methodology}
\end{figure}

The framework we developed for RAM demand modeling, illustrated in \autoref{fig:Methodology}, integrates various essential components: generating trip demand from available datasets, developing regression model associated risks, costs, travel time, and GTC. Finally predicting the demand for RAM. The trip demand data from the Next-Generation National Household Travel Survey (NextGen NHTS) is specified at the MSA level\cite{FHWA2022}, it is presumed that travel originates from the population centroid of the departure MSA and ends at that of the arrival MSA. MSAs are defined by the Office of Management and Budget (OMB) and are used by federal statistical agencies for collecting, analyzing, and publishing statistical data about specific geographic regions. An MSA is essentially a region that consists of one or more counties that contain an urbanized area of 50,000 or more inhabitants. Prior to December 2003,Tennessee had seven MSAs. Under the new guidelines in December, three more were added, bringing the total to 10 MSAs\cite{Helmer2008MicropolitanSA}. Detail about these MSAs are presented in the \autoref{tab:MSA}. Counties in the east which don't fall under MSAs are grouped as TN-NonMSA areas(E) and similarly for the west are grouped as TN-NonMSA areas(W) with CBSA code as RTN1 and RTN3 respectively\footnote{Planned Passenger Travel Origin Destination Zone Information (Federal Highway Administration): \url{ https://www.fhwa.dot.gov/policyinformation/analysisframework/04.cfm }.Last accessed on [May 9\textsuperscript{th} 2024]}. We collected the monthly passenger Origin-Destination(OD) data for two year 2021 and 2022. First we analyzed all the trips data, and later we filtered out the trip that satisfy the RAM requirement which is range of the distance should be from 50 to 500 miles. A brief summary of the dataset is given in the \autoref{tab:Analysis}.

We have the following assumptions for the trips:
\begin{enumerate}
    \item All trips generated from MSAs are generated from the centroid of population of MSAs.
    \item Ground transportation trip consists of distance travel from centroid of origin MSA to centroid of destination MSA.
    \item RAM trip consists of ground transportation from centroid of origin MSA to the nearest hub airport and then RAM flight to destination airports and finally ground transportation from destination airports to the centroid of destination MSA.
\end{enumerate}

\begin{table}
\caption{\label{tab:Analysis} Comparison of Trip Demand Features}
\centering
\begin{tabular}{ccc}
\hline
\textbf{Feature} & \textbf{All Trips} & \textbf{Trips (50–500 miles)} \\ \hline
\textbf{Trip Distance Distribution} & \makecell[l]{0-10 miles: 72.67\% \\ 10-25 miles: 20.04\% \\ 25-50 miles: 5.46\% \\ 50-75 miles: 0.85\% \\ 75-100 miles: 0.38\% \\ 100-150 miles: 0.32\% \\ 150-300 miles: 0.25\% \\ $>$300 miles: 0.03\% }
&
\makecell[l]{50-75 miles: 46.31\% \\ 75-100 miles: 20.71\% \\ 100-150 miles: 17.51\% \\ 150-300 miles: 13.68\% \\ $>$300 miles: 1.79\% }
\\ \hline

\textbf{Preferred Mode of Transportation} & \makecell[l]{Vehicle: 92.609\% \\ Active Transportation/Ferries: 7.380\% \\ Rail: 0.009\% \\ Airways:0.002\%}
& 
\makecell[l]{Vehicle: 99.892\% \\ Airways:0.095\% \\ Rail: 0.013\% \\Active Transportation/Ferries: 0.0\%  } \\ \hline

\textbf{Purpose of Trip} & \makecell[l]{Work: 25.81\% \\ Non\_Work: 74.19\%} 
& 
\makecell[l]{Work: 29.46\% \\ Non\_Work: 70.54\%} \\ \hline

\textbf{Monthly Trip Demand} & \makecell[l]{Highest: August \\ Lowest: February}
& 
\makecell[l]{Highest: October, followed by August \\ Lowest: February}
\\ \hline

\textbf{Trip Demand by MSA} & \makecell[l]{Highest: 34980 (29.45\%) \\ Lowest: 34100 (1.78\%)}
& 

\makecell[l]{Highest: 34980 (30.41\%) \\ Lowest: 17420 (1.30\%)} 
\\ \hline

\textbf{Most Popular OD Pair} & \makecell[l]{Highest: RTN1 to 28940 (8.59\%) \\ Lowest: 27180 to 27740  (0.0006\%)} 
& 
\makecell[l]{Highest: RTN3 to 34980 (8.35\%) \\ Lowest: 27180 to 27740 (0.002\%)}
\\ \hline
\end{tabular}
\end{table}



\begin{table}
\caption{\label{tab:MSA} Metropolitan Statistical Areas in Tennessee State}
\centering
\begin{tabular}{ccc}
\hline
\textbf{Metropolitan Statistical Areas}& \textbf{CBSA code} & \textbf{Counties Included}\\ \hline

 Chattanooga, TN-GA & 16860 & \makecell[l]{Hamilton, Marion, Sequatchie,
Catoosa(GA),\\ Dade(GA), Walker(GA)}\\ \hline

Clarksville, TN-KY & 17300& \makecell[l]{Montgomery, Christian(KY), Trigg(KY)}\\ \hline

 Cleveland, TN & 17420&\makecell[l]{Bradley, Polk}\\
\hline
 
 Jackson, TN &27180& \makecell[l]{Chester, Madison, Crockett}\\ \hline
 
 Johnson City, TN &27740& \makecell[l]{Carter, Unicoi, Washington}\\ \hline

 Kingsport-Bristol-Bristol, TN-VA&28700& \makecell[l]{Hawkins, Sullivan,  Scott(VA), Washington(VA),\\ Bristol(VA)} \\ \hline

 Knoxville, TN &28940& \makecell[l]{Anderson, Blount,Campbell, Grainger, Knox, \\ Loudon,Morgan, Roane, Union} \\ \hline

 Memphis, TN-MS-AR &32820& \makecell[l]{Fayette, Shelby, Tipton, DeSoto(MS),Tate(MS), \\ Marshall(MS), Tunica(MS),Benton(MS),\\ Crittenden(AR)} \\ \hline

 Morristown, TN &34100& \makecell[l]{ Hamblen, Jefferson}\\ \hline

 Nashville-Davidson-Murfreesboro-Franklin, TN &34980& \makecell[l]{ Cannon, Maury,
Cheatham,
Davidson,
Dickson,\\
Hickman,
Macon, 
Robertson,
Rutherford,
Smith,\\
Sumner,
Trousdale,
Williamson,
Wilson} \\

\hline
\end{tabular}
\end{table}

Five major airports in Tennessee state are chosen as the hub for the RAM transportation. They are chosen on the basis of their location which are near to the densely populated MSAs and well distributed geographical position. The details about those airports are given in \autoref{tab:Airports}.

\begin{figure}[hbt!]
\centering
\includegraphics[width=\textwidth]{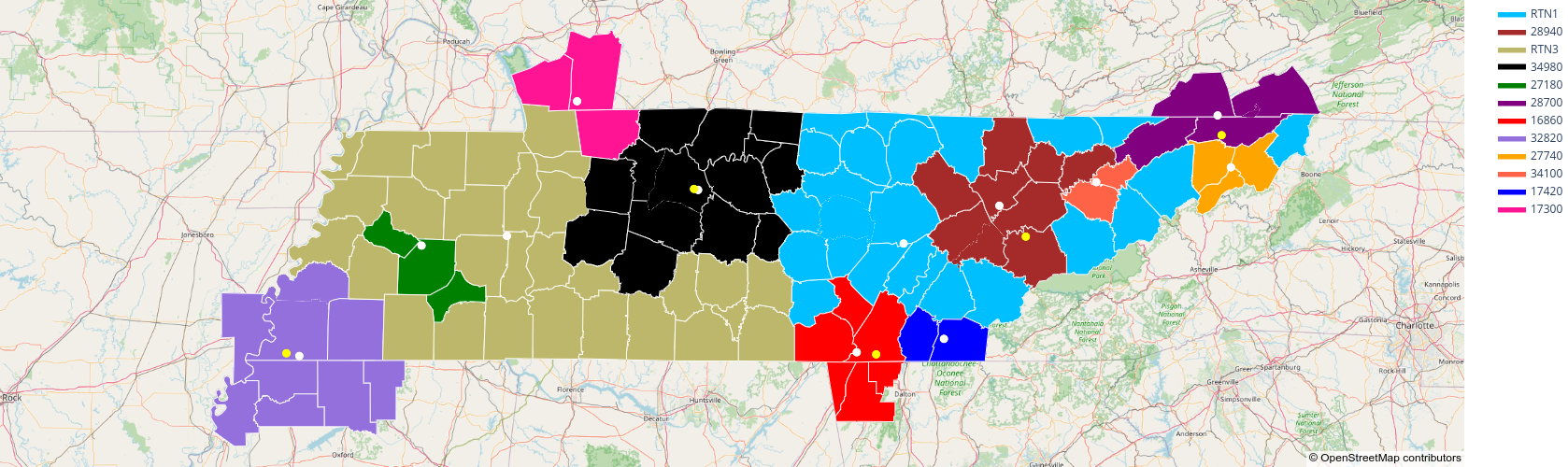}
\caption{Tennessee State Region for Implementation}
\label{fig:MSA}
\end{figure}

We mapped the 10 MSAs including RTN1 and RTN3 into one of the 5 airports to develop them as the hub for the RAM for the trip demand generated from each of these locations. \autoref{fig:MSA} shows the detail about each MSAs along with the centroid of MSAs\footnote{Gazetteer Files: \url{ https://www.census.gov/geographies/reference-files/time-series/geo/gazetteer-files.2021.html\#list-tab-264479560 }.Last accessed on [May 9\textsuperscript{th} 2024]} represented by white circle and location of the airports under consideration by yellow circle.

\subsection{Dataset Used}
\autoref{tab:Datasets} summarizes the various datasets used in this research. The Bureau of Transportation Statistics (BTS) Monthly Traffic Dataset supplies data on transportation-related fatalities, allowing for risk assessment, while the U.S. Department of Transportation (USDoT) Value of Statistical Life (VSL)\cite{USDOT2024}  Dataset quantifies the monetary value of reducing mortality, supporting the integration of safety into cost modeling. To gauge trip demand, the National Household Travel Survey (NHTS) Dataset offers demand data at the MSA level, and Gazetteer Files provide geographic centroids for MSAs, ensuring accurate origin-destination mappings. For cost estimates, the BTS Data Bank 1B (DB1B) Market Dataset gives ticket prices for airlines, while the IRS Standard Mileage Dataset provides mileage rates essential for calculating ground travel costs. Time estimates derive from the FAA Airport Dataset on block times for flights and the BTS Inter-Airport Distance Dataset for calculating air travel distances. Google Maps API data contribute ground transportation distances and times, while the U.S. Bureau of Labor Statistics dataset provides median hourly wages, essential for estimating the economic value of time in the GTC model.

For ground transportation, distances are calculated as driving distances obtained via the Google Maps Distance Matrix API\footnote{Distance Matrix API: \url{ https://developers.google.com/maps/documentation/distance-matrix }. Accessed on [[May 9\textsuperscript{th} 2024]]} whereas for air travel, distances are measured as great-circle distances, computed using the cosine haversine formula\cite{soe2020haversine}, detailed below.


\begin{table}
\caption{\label{tab:Datasets} Datasets Used in the Research}
\centering
\begin{tabular}{cc}
\hline
\textbf{Name }& \textbf{Details}\\ \hline

 BTS Monthly Traffic Dataset & Number of fatalities during Transportation \\ \hline

  USDoT VSL Dataset & Monetary equivalent of reducing one death in population \\ \hline

NHTS Dataset & Trip Demand in MSA level\\ \hline

 Gazetteer Files& Centroid of population of MSA\\
\hline
 
 BTS DB1BMarket Dataset & Ticket Price for the airlines\\ \hline

 IRS Standard Mileage Dataset & Standard Mileage Rates(cents/miles) \\ \hline
 
 FAA Airport Dataset& Block time of the flights\\
\hline
 BTS Inter-Airport Distance Dataset & Distance between the airports\\
\hline

 Google Map API & Distance and Time of Ground transportation\\
 \hline

 US Bureau of Labor Statistics & Median hourly wages\\
\hline
\end{tabular}
\end{table}

\subsection{Cost Regression}
The cost regression assesses the total expenses associated with complete door-to-door trips, encompassing various journey segments that involve multiple modes of transportation, including travel to and from airports.

The costs of driving are calculated using the standard mileage rates provided by the United States Internal Revenue Services\footnote{Internal Revenue Services (IRS): \url{ https://www.irs.gov/tax-professionals/standard-mileage-rates }.Last accessed on [May 9\textsuperscript{th} 2024]}, coupled with driving distance estimates derived from the Google Map Distance Matrix API Service. 

\begin{equation}
\label{equation:CostGroundTransportation}
C_G= d * \beta
\end{equation}

The costs associated with air travel are influenced by multiple factors, such as travel distance, service class, booking lead time, time of day, day of the week, market competition, and the level of airline concentration at the origin or destination. This study employs a simplified model that links passenger costs directly to distance. The data for this model is drawn from the Department of Transportation's (DOT) DB1B database, which provides a 10\% sample of all tickets sold in the United States during 2021 and 2022. This database consists of three segments: ticket, market, and coupon data. Among them, the market database is utilized to regress the airfare.

\begin{table}
\caption{\label{tab:Airports} Major Airports in Tennessee under consideration for RAM station}
\centering
\begin{tabular}{ccc}
\hline
\textbf{Airports}& \textbf{IATA Code} & \textbf{Location(MSA)}\\ \hline

 Nashville International Airport & BNA
 & Nashville (34980) \\ \hline

Memphis International Airport& MEM
 & Memphis (32820)\\ \hline

 McGhee Tyson Airport& TYS
 & Alcoa (28940)\\
\hline
 
 Lovell Field Airport & CHA
 & Chattanooga (16860)\\ \hline
 
 Tri-Cities Airport & TRI & Blountville (28700)\\

\hline
\end{tabular}
\end{table}

\subsection{Time Regression}

Driving times are calculated using Google Maps' Distance Matrix API, capturing data for both off-peak (no traffic) and peak (with traffic) hours, with the origin and destination points represented by the population centroids of the respective Metropolitan Statistical Areas (MSAs). Flight travel times are estimated using block time, which measures the interval from when an aircraft’s brakes are released at departure to when they are engaged upon arrival. This information is derived from the 2021 and 2022 Aviation System Performance Metrics (ASPM) dataset\footnote{FAA Aviation System Performance Metrics (ASPM): \url{https://aspm.faa.gov/apm/sys/AnalysisCP.asp}. Last accessed on [May 9\textsuperscript{th}, 2024]}, accessible via the FAA’s online platform.

\subsection{Risk Regression}

 In this study, we calculate the fatality rates associated with each mode of transportation. The Value of Statistical Life (VSL), an economic metric used to measure the societal benefit of reducing death risk, represents the monetary amount society is willing to invest to save a life \cite{de2003value}. By applying VSL to fatality estimates, we derive a monetary equivalent for the risks associated with each transportation mode, calculated as follows:
 \begin{equation}
\label{equation:Risk}
\begin{aligned}
R_G= VSL * \alpha_G \\
R_A= VSL * \alpha_A
\end{aligned} 
\end{equation}


In this study, we utilized the fatality rate per 100 million vehicle miles traveled for ground transportation from the National Safety Council (NSC) dataset\footnote{Motor-Vehicle Deaths by State:\url{ https://injuryfacts.nsc.org/state-data/motor-vehicle-deaths-by-state/ }. Accessed on [May 9\textsuperscript{th} 2024]}. Similarly, we calculated the air travel fatality rate per million miles using data on airplane crashes from the NSC\footnote{Airplane Crashes:\url{ https://injuryfacts.nsc.org/home-and-community/safety-topics/airplane-crashes/ }. Accessed on [May 9\textsuperscript{th} 2024]}.

\subsection{Generalized Travel Cost Regression}

Researchers often utilize disaggregate models to predict mode choice, where the utility of each transport option is assessed based on factors unique to both the traveler and the mode. However, building utility functions that incorporate numerous attributes can be intricate and typically demands extensive data collection from surveys. To address this complexity, \cite{tanner1981expenditure} introduced the concept of generalized cost. The generalized cost consists of two main components: the direct cost and the opportunity cost associated with the trip. The direct cost includes immediate expenses, such as fuel or ticket prices, while the opportunity cost captures the economic worth of time spent traveling, based on median hourly wage estimates\footnote{U.S. Bureau of Labor Statistics: \url{https://www.bls.gov/oes/tables.htm}. Accessed on [May 9\textsuperscript{th} 2024]} for the MSAs at both the origin and destination. In the GTC calculation, each negative term represents the “disutility” or drawbacks associated with increased expenses, longer travel times, and additional risks, effectively modeling traveler decision-making.

\begin{equation}
\label{equation:GTC}
\begin{aligned}
GTC_G= -C_G - W*T_G - R_G \\
GTC_A= -C_A - W*T_A - R_A
\end{aligned} 
\end{equation}

The GTC for a RAM trip is determined by summing the GTC of the corresponding ground and air transportation segments.

\begin{equation}
\label{equation:GTC_RAM}
GTC_{RAM} = GTC_G + GTC_A
\end{equation}

\subsection{Probability of Selection}
The GTC metric works well in models that presume individuals make choices with complete rationality and consistency. However, this method may miss subtle factors that influence personal choices. To incorporate these nuances, a probability-based model is often more effective, as it evaluates the likelihood of selecting a particular transportation option instead of assuming a specific decision. In this model, the overall benefit of a choice is represented by a structured component combined with an unpredictable variation or error, offering a more adaptable and authentic perspective on travel behavior.

\begin{equation}
\label{equation:Utility}
\begin{aligned}
U_{\text{G}} &=  GTC_{\text{G}} + \epsilon \\
U_{\text{RAM}} &=  GTC_{\text{RAM}} + \epsilon
\end{aligned} 
\end{equation}

Traveler behavior in choosing between transportation modes is often represented using discrete choice models, which calculate the likelihood of opting for a particular mode. Examples of these models include the logit and conditional logit models, distinguished by their assumptions regarding the distribution of the error term (\(\epsilon\)) in the utility equation. A typical assumption is that this error term is independently and identically distributed (IID), which leads to the multinomial logit model\cite{heiss2016discrete}. This model provides a straightforward formula for determining choice probabilities based on the utility differences among transportation options. By incorporating the unpredictability in individual preferences through the error term, discrete choice models create a probability-based structure that mirrors real-world choices more closely. Consequently, the probability of selecting a specific mode can be formulated as follows: 
\begin{equation}
\label{equation:Probability}
\begin{aligned}
P_{\text{RAM}} = \frac{1}{1 + e^{(U_{\text{G}} - U_{\text{RAM}})}} \\
P_{\text{RAM}} = \frac{1}{1 + e^{ (GTC_{\text{G}} - GTC_{\text{RAM}})}}
\end{aligned}
\end{equation}

\section{Results}
\begin{figure}[hbt!]
    \centering
    \subfloat[MSE with various degrees of Polynomial]{
        \includegraphics[width=0.42\textwidth]{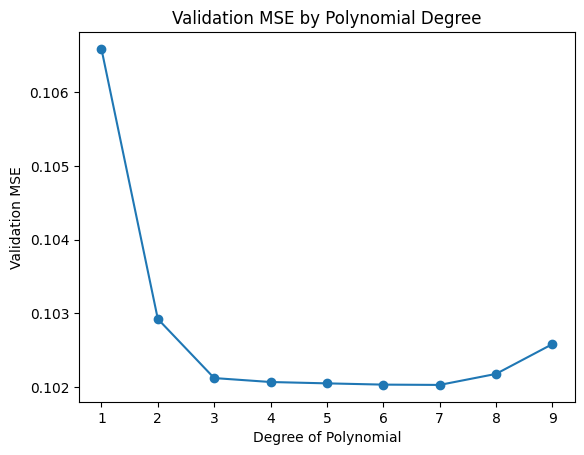}
        \label{fig:CostDegree}
    }
    \hfill
    \subfloat[Cost Regression]{
        \includegraphics[width=0.45\textwidth]{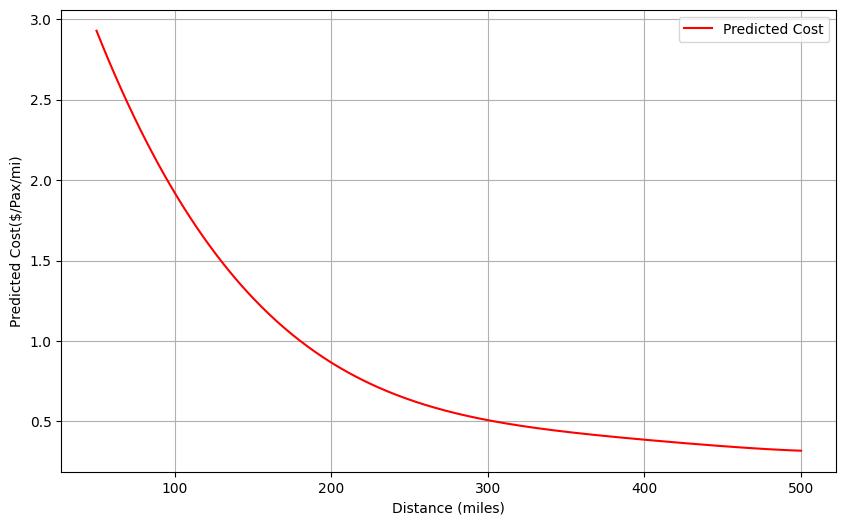}
        \label{fig:CostRegression}
    }
    \caption{Cost Regression Analysis for Air Transportation}
    \label{fig:Cost}
\end{figure}

The regression model for cost optimization identified a polynomial degree of 4, as shown in Figure \autoref{fig:CostDegree}. This model was utilized to estimate prices for various distances ranging from 50 to 500 miles. The analysis, visualized in Figure \autoref{fig:CostRegression}, demonstrates the variation in passenger cost per mile with travel distance for flights in Tennessee. The results reveal an inverse relationship: shorter flights tend to have a higher cost per mile compared to longer ones. This pattern aligns with the economies of scale in air travel, where fixed costs are spread over longer distances, reducing the per-mile cost on extended routes.

\begin{figure}[hbt!]
    \centering
    \subfloat[MSE with various degrees of Polynomial]{
        \includegraphics[width=0.4\textwidth]{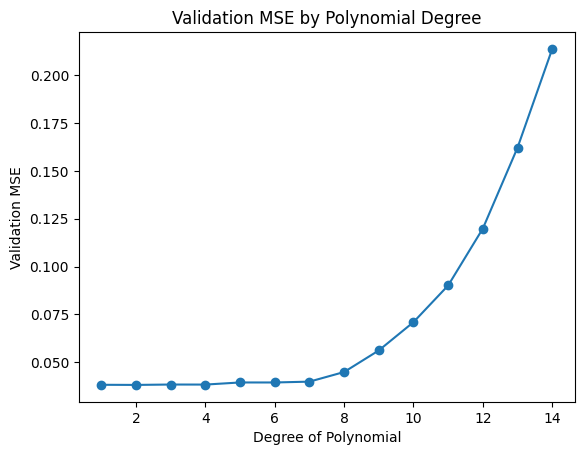}
        \label{fig:TimeDegree}
    }
    \hfill
    \subfloat[Time Regression Analysis For Ground Transportation]{
        \includegraphics[width=0.45\textwidth]{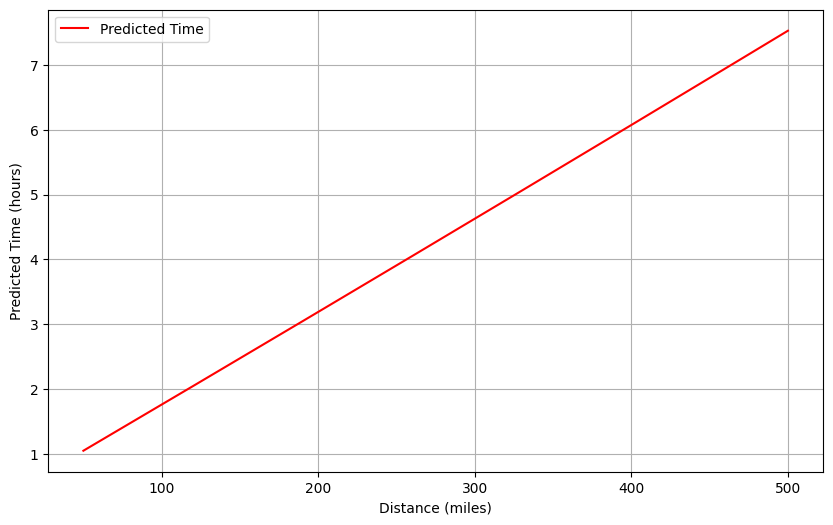}
        \label{fig:Time}
    }
    \caption{Time Regression Analysis for Ground Transportation}
    \label{fig:TimeRegressionGround}
\end{figure}

The validation MSE for ground time regression across different polynomial degrees is presented in Figure \autoref{fig:TimeDegree}. The results reveal that lower polynomial degrees yield higher accuracy, leading to the selection of degree 1 for our regression model. Furthermore, the regression analysis shown in Figure \autoref{fig:Time} demonstrates a strong positive correlation between travel time and distance, indicating that longer trips typically require more time.

\begin{figure}[hbt!]
    \centering
    \subfloat[MSE with various degrees of Polynomial]{
        \includegraphics[width=0.4\textwidth]{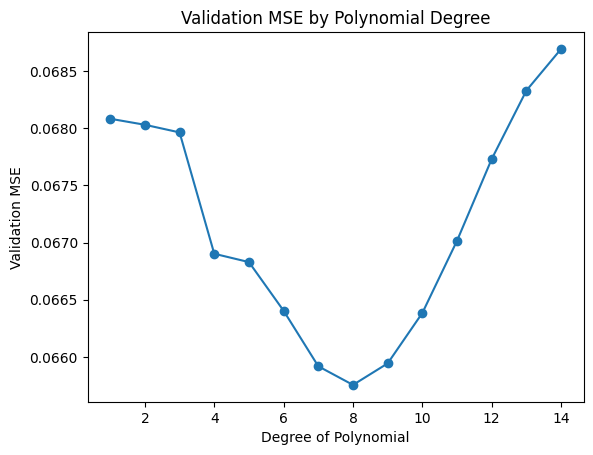}
        \label{fig:BlockTimeDegree}
    }
    \hfill
    \subfloat[Block Time Regression Analysis For Airlines]{
        \includegraphics[width=0.45\textwidth]{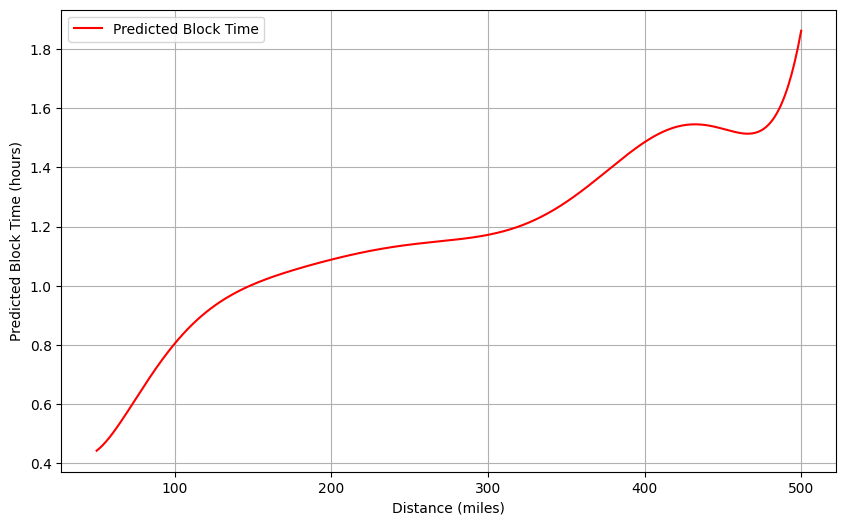}
        \label{fig:BlockTime}
    }
    \caption{Time Regression Analysis for Air Transportation}
    \label{fig:TimeRegressionAirlines}
\end{figure}

In aviation, block times tend to follow a linear or polynomial trend, influenced by factors like air traffic volume, airport congestion, and operational delays. Considering this pattern, we adopt a polynomial fit in our time modeling to better capture these dynamics. The Figure \autoref{fig:BlockTimeDegree} depicts the validation MSE across various polynomial degrees, identifying the 8th-degree polynomial as the optimal model for capturing the relationship. The regression analysis in Figure \autoref{fig:BlockTime} illustrates the fitted regression line, revealing a positive correlation between travel distance and block time. The analysis highlights substantial variability, particularly for shorter distances, likely influenced by factors such as airport congestion, layovers, and scheduling irregularities. The regression line demonstrates a nonlinear increase in block time with distance, reflecting efficiency improvements on longer flights. Notably, at distances around 480 miles, block times begin to rise more steeply, possibly due to an increase in layovers that extend time spent at connecting airports.

\begin{figure}[hbt!]
\centering
\includegraphics[width=0.5\textwidth]{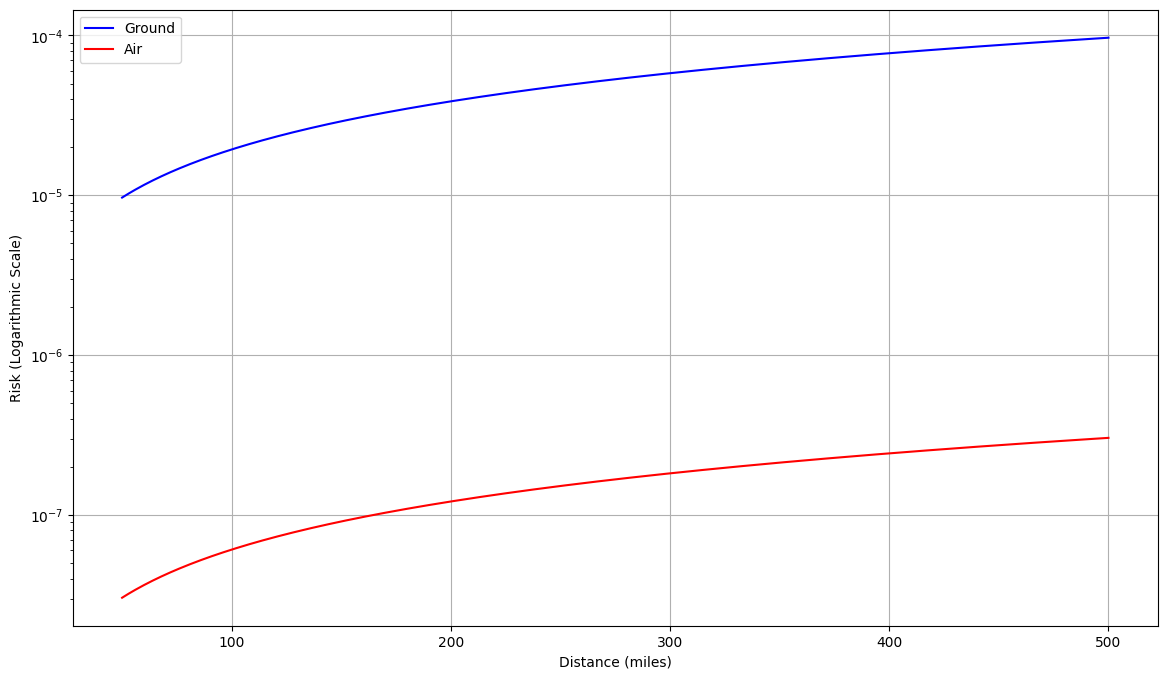}
\caption{Risk regression for different modes of Transportation}
\label{fig:risk}
\end{figure}

The analysis presented in \autoref{fig:risk} demonstrates that, on a logarithmic scale, ground transportation poses a higher fatality risk cost than air travel. This underscores the relative safety of air travel, even though both modes show an increase in risk with greater distances. Nevertheless, the financial implications of these risks remain insignificant when compared to other cost involved.

\begin{figure}[hbt!]
\centering
\includegraphics[width=0.5\textwidth]{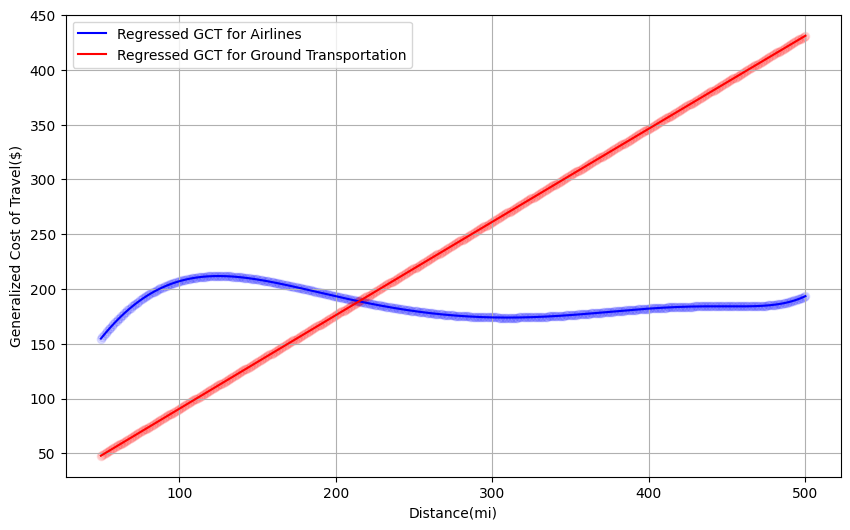}
\caption{Regression Analysis For Generalized Travel Cost}
\label{fig:GTC}
\end{figure}

The regression analysis in \autoref{fig:GTC} shows that the GTC for ground transportation increases linearly with distance, indicating a consistent rise in costs as travel distance extends. In contrast, the GTC for airlines exhibits a non-linear pattern, initially increasing, then slightly decreasing, and finally stabilizing, suggesting that airline travel may become more cost-efficient over certain distances. This pattern likely reflects the fixed costs and economies of scale associated with air travel, which become more pronounced over longer distances. Notably, for distances around 480 miles, the GTC for airlines increases again, possibly due to additional costs associated with layovers and transit times at airports for longer journeys.

\subsection{Generalized Travel Cost Modeling for RAM}
\subsubsection{CASE I: Considering only the five major airports}

\begin{figure}[hbt!]
\centering
\includegraphics[width=0.5\textwidth]{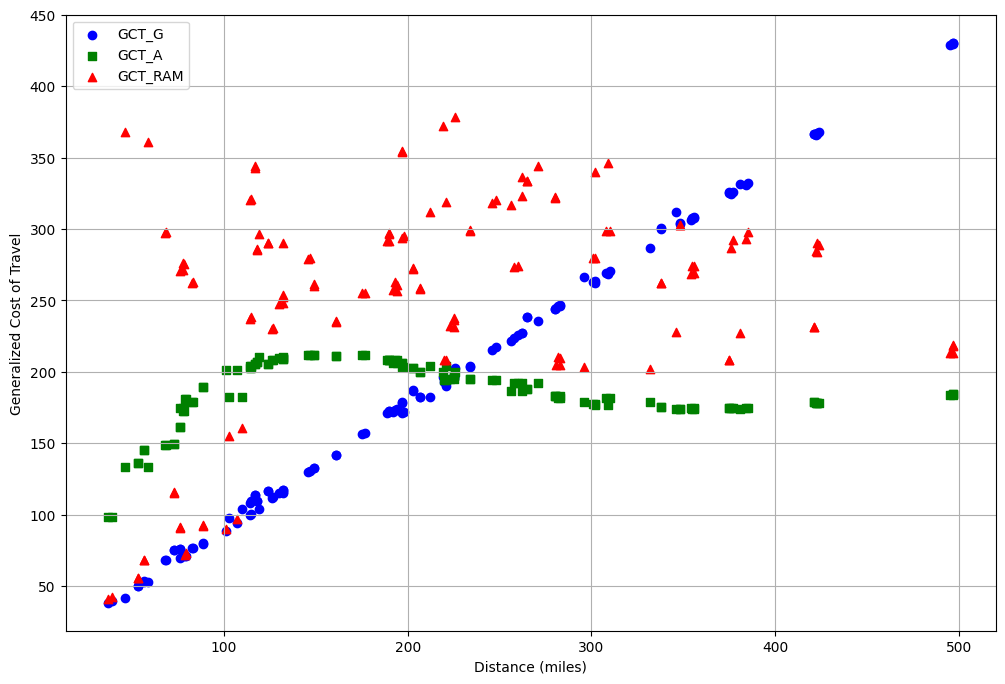}
\caption{Generalized Travel Cost for various modes for different OD pairs for Case I}
\label{fig:GTC_all}
\end{figure}

\autoref{fig:GTC_all} illustrates the GTC for all Origin-Destination (OD) pairs of MSA. The GTC for airlines and ground transportation is self-explanatory, as it is significantly influenced by the cost models and exhibits similar patterns. The GTC for RAM, however, is highly variable due to several reasons outlined below:
\begin{enumerate}
    \item Since only five commercial airports are considered, multiple MSA are mapped to a single airport, resulting in the following scenarios:
    \begin{itemize}
        \item If the destination MSA is in the same direction of the airport, it is almost equivalent to taking ground transportation, and the point appears close to the ground transportation values.
        \item If the destination is in a different direction than the airport, there is back-and-forth movement, causing the GTC of RAM to be significantly higher compared to ground transportation.
    \end{itemize}
    \item When MSAs are mapped to different airports for longer distances, generally exceeding 250 miles, the GTC of RAM tends to be lower than that of ground transportation.
\end{enumerate}

The above scenario clearly indicates the necessity of adding more airports as stations for RAM transportation because of irregular patterns obtained for RAM transportation.
\subsubsection{CASE II: Considering all the airports}
\begin{figure}[hbt!]
\centering
\includegraphics[width=\textwidth]{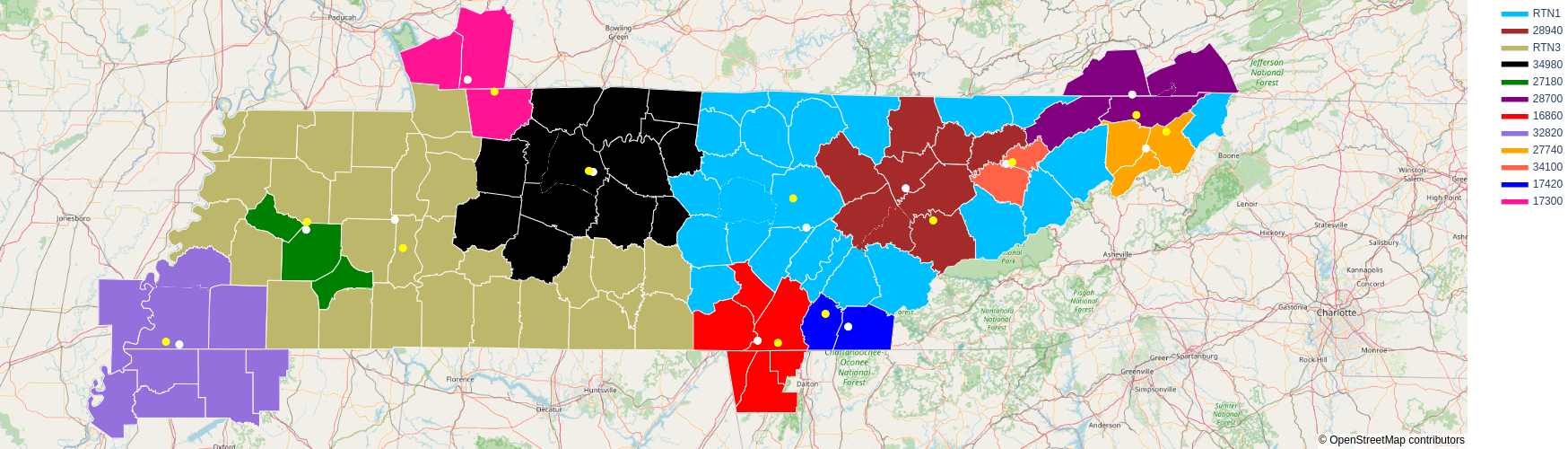}
\caption{MSAs with corresponding nearest airports in Tennessee State Region}
\label{fig:MSA_more_airports}
\end{figure}

All airports in the state of Tennessee are considered, and the nearest airports to the centroids of MSA are identified and designated as RAM ports for those MSA. The details are provided in \autoref{tab:AirportsMapping} and visualized in \autoref{fig:MSA_more_airports}.

\begin{table}
\caption{\label{tab:AirportsMapping} Nearest Airports from MSA}
\centering
\begin{tabular}{ccc}
\hline
\textbf{Airports}& \textbf{IATA Code} & \textbf{MSA (City)}\\ \hline

 Nashville International Airport & BNA
 & 34980 (Nashville) \\ \hline

Memphis International Airport& MEM
 & 32820 (Memphis)\\ \hline

 McGhee Tyson Airport& TYS
 & 28940 (Alcoa)\\
\hline
 
 Lovell Field Airport & CHA
 &16860 (Chattanooga\\ \hline
 
 Tri-Cities Airport& TRI & 28700 (Blountville)\\ \hline

 Elizabethton Municipal &0A9	&	27740 (Elizabethton)\\ \hline

 Scott & 0M1	&	RTN3 (Parsons)	\\ \hline

 Outlaw &CKV	 &	17300 (Clarksville)		\\ \hline
 
 Crossville Memorial&CSV	&	RTN1 (Crossville)		\\ \hline

 Hardwick &HDI	&	17420 (Cleveland)	\\ \hline

 Humboldt Municipal &M53	&	27180 (Humboldt) \\ \hline

 Moore-Murrell  & MOR	&	34100 (Morristown) \\

\hline
\end{tabular}
\end{table}

\begin{figure}[hbt!]
\centering
\includegraphics[width=0.5\textwidth]{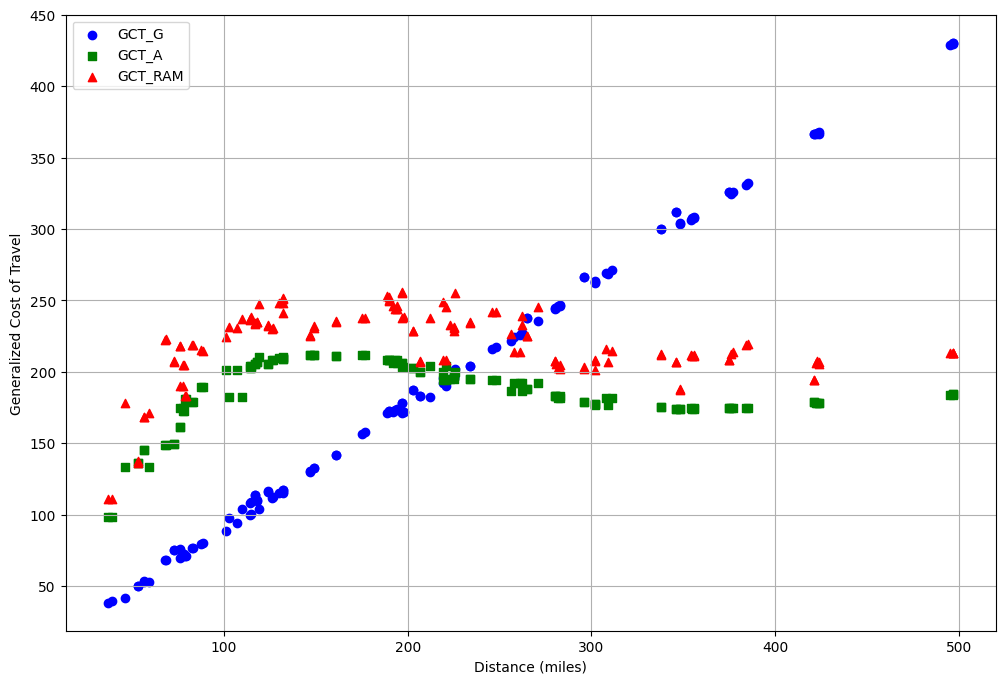}
\caption{GTC for various modes for different OD pairs  for Case II}
\label{fig:GTC_more_airports}
\end{figure}

The \autoref{fig:GTC_more_airports} displays the GTC for all modes of transfer. It can be observed that the GTC of RAM closely follows the pattern of the GTC of airlines and is lower than the GTC of ground transportation for distances of greater than around 260 miles.

\subsection{Probability of Selection}
\begin{figure}[hbt!]
    \centering
    \subfloat[Probability of selecting transportation mode]{
        \includegraphics[width=0.4\textwidth]{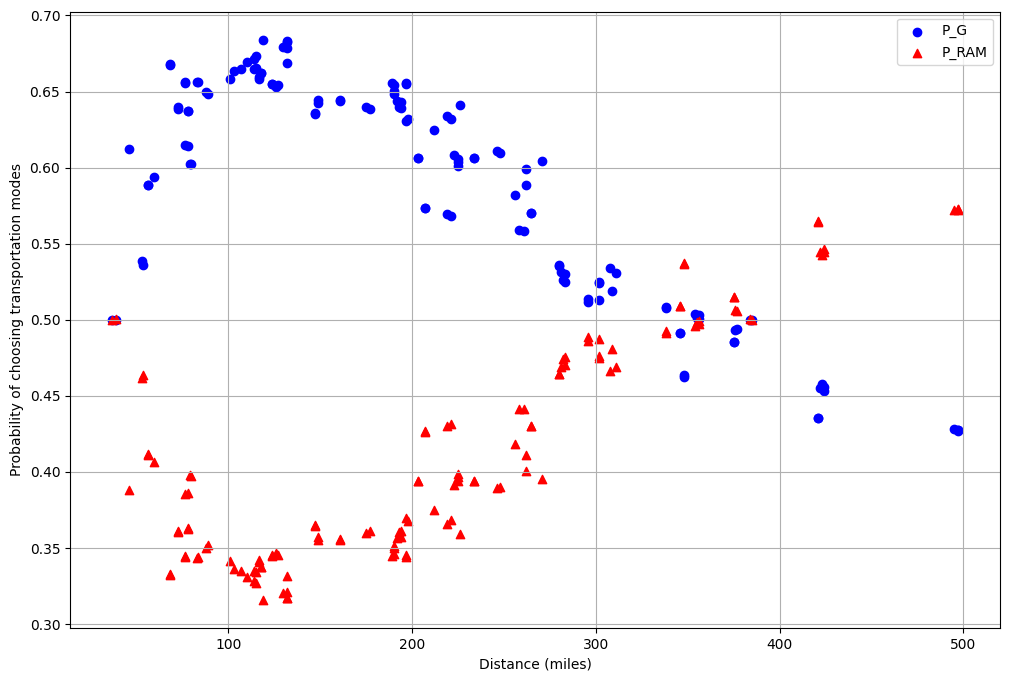}
        \label{fig:modechoiceDistance}
    }
    \hfill
    \subfloat[Percentage of $GTC_A$ in $GTC_{RAM}$ vs difference of $P_G$ and $P_{RAM}$]{
        \includegraphics[width=0.4\textwidth]{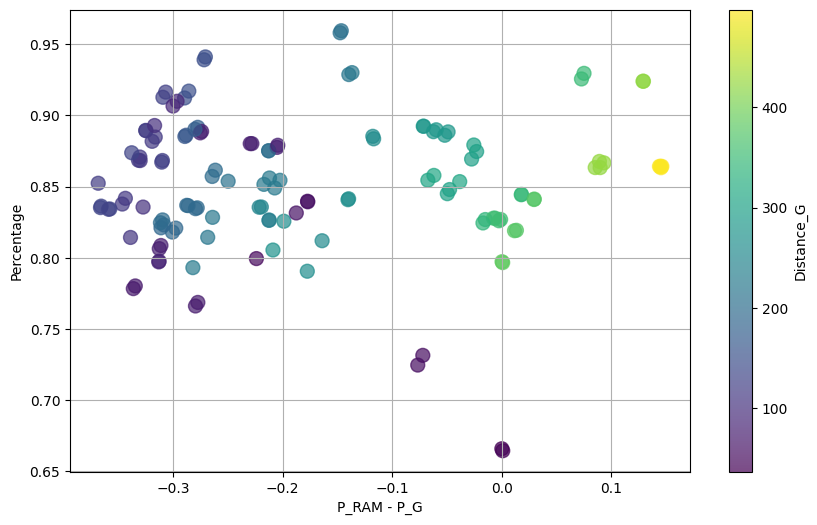}
        \label{fig:modechoicePercentage}
    }
    \caption{Probability of Selection Analysis}
    \label{fig:choiceMode}
\end{figure}

For the CASE II, we performed the calculation for finding the probability of choosing the mode of transportation between RAM and ground transportation using equation (5). The results obtained are shown in \autoref{fig:choiceMode}.

The Figure\autoref{fig:modechoiceDistance} illustrates the probability of choosing different transportation modes as distance (in miles) increases. For shorter trips, ground transportation is preferred, maintaining a higher probability initially. However, this likelihood declines with increasing distance, reflecting a shift toward RAM for longer journeys. Specifically, ground transportation is chosen more frequently for distances up to around 200 miles, but regional air mobility becomes the preferred option for trips exceeding 300 miles. This trend suggests a distance threshold where regional air mobility becomes more appealing than ground travel for extended trips. The Figure\autoref{fig:modechoicePercentage} demonstrates that, beyond the threshold distance, demand for regional air mobility is also greater for trips where the air component of the GTC exceeds 80\% of the total GTC, highlighting its attractiveness for longer-distance, high-cost trips.

\begin{figure}[hbt!]
    \centering
    \subfloat[Origin-Destination Pairs Switched to RAM]{
        \includegraphics[width=0.48\textwidth]{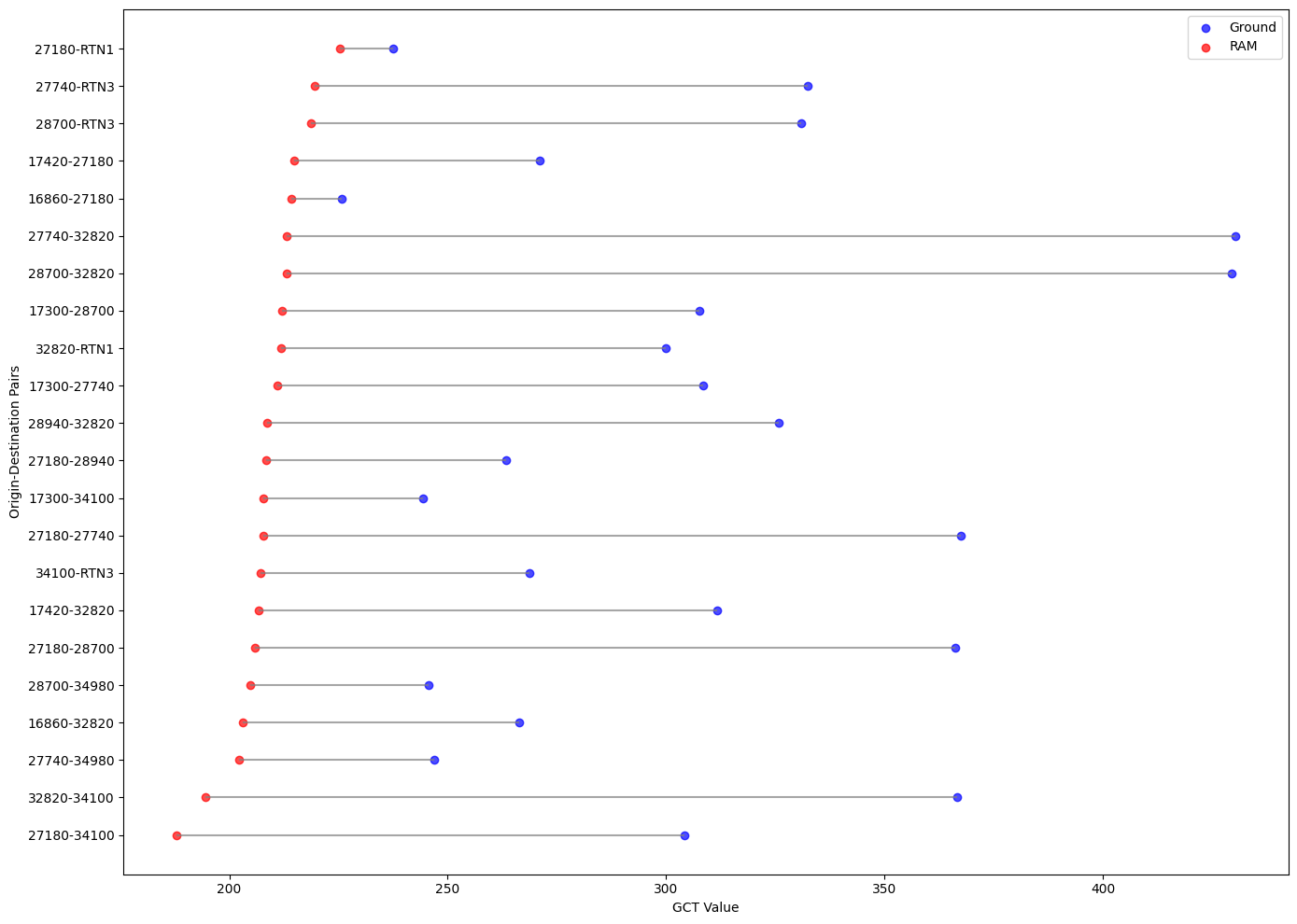}
        \label{fig:switched}
    }
    \hfill
    \subfloat[Origin-Destination Pairs Unswitched to RAM]{
        \includegraphics[width=0.48\textwidth]{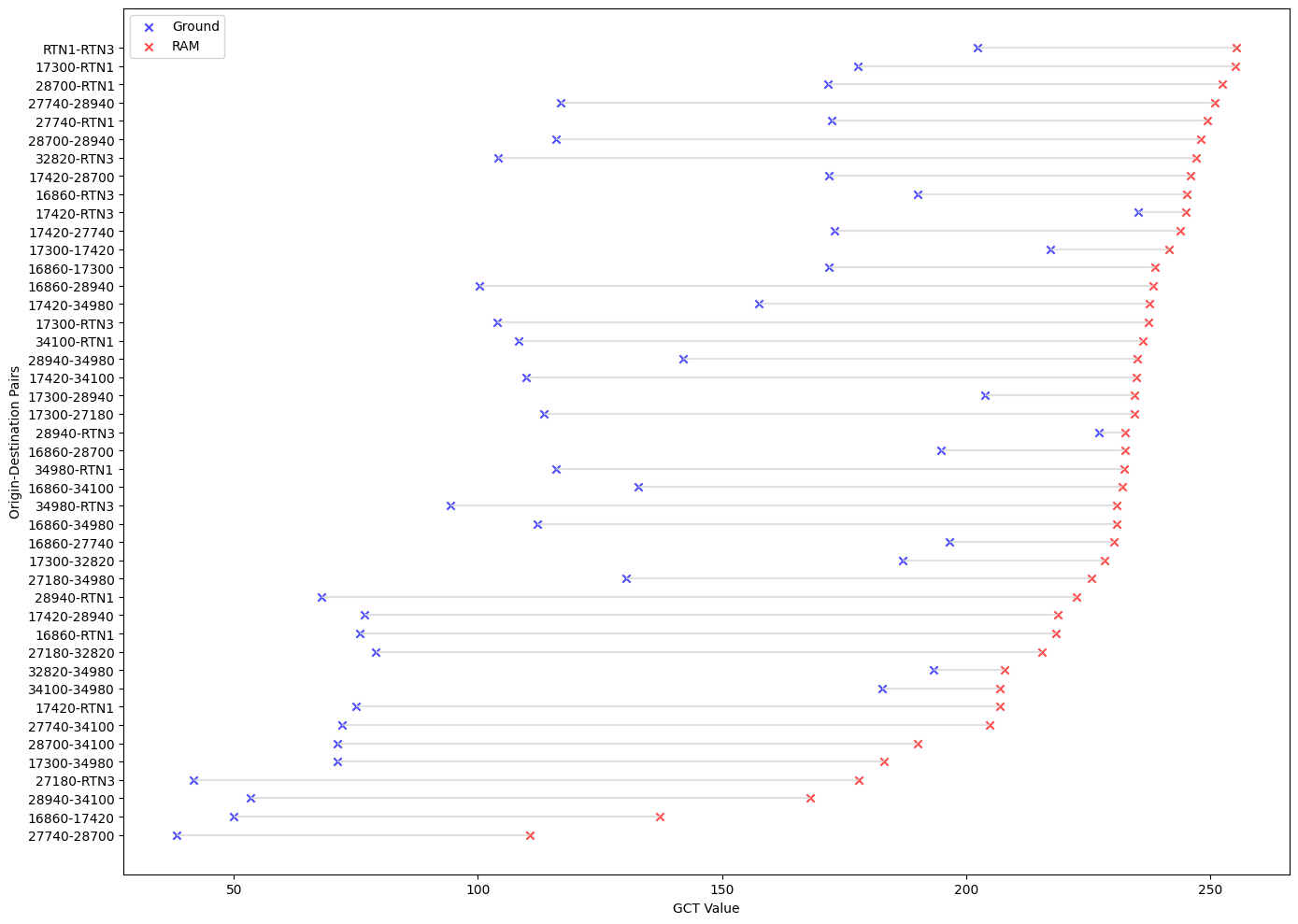}
        \label{fig:unswitched}
    }
    \caption{Switching of OD pairs}
    \label{fig:switching}
\end{figure}

The \autoref{fig:switching} provides a comparative analysis of OD pairs in terms of their GTC and mode switching behavior. The Figure \autoref{fig:switched} highlights the OD pairs that transitioned from ground transportation to RAM, while Figure \autoref{fig:unswitched} showcases those that remain unswitched. The data reveals a distinct trend: OD pairs with relatively higher GTC values for ground transportation tend to switch to RAM, with a stabilization range of GTCs for RAM observed around $ \$ 200-250$ . Conversely, OD pairs with lower GTC values for ground transportation exhibit a stronger preference for retaining the ground mode, indicating cost efficiency as a key determinant in transportation mode selection.

\begin{figure}[hbt!]
    \centering
    \subfloat[RAM Trip]{
        \includegraphics[width=0.48\textwidth]{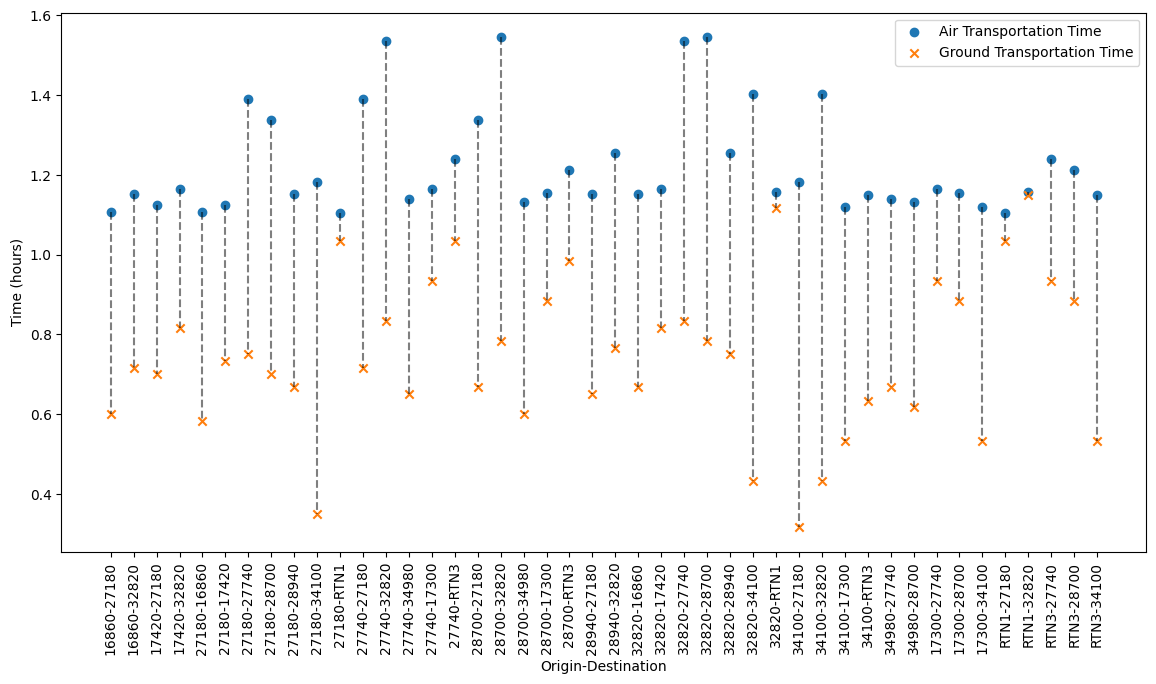}
        \label{fig:RAMTripTime}
    }
    \hfill
    \subfloat[Non-RAM Trip]{
        \includegraphics[width=0.48\textwidth]{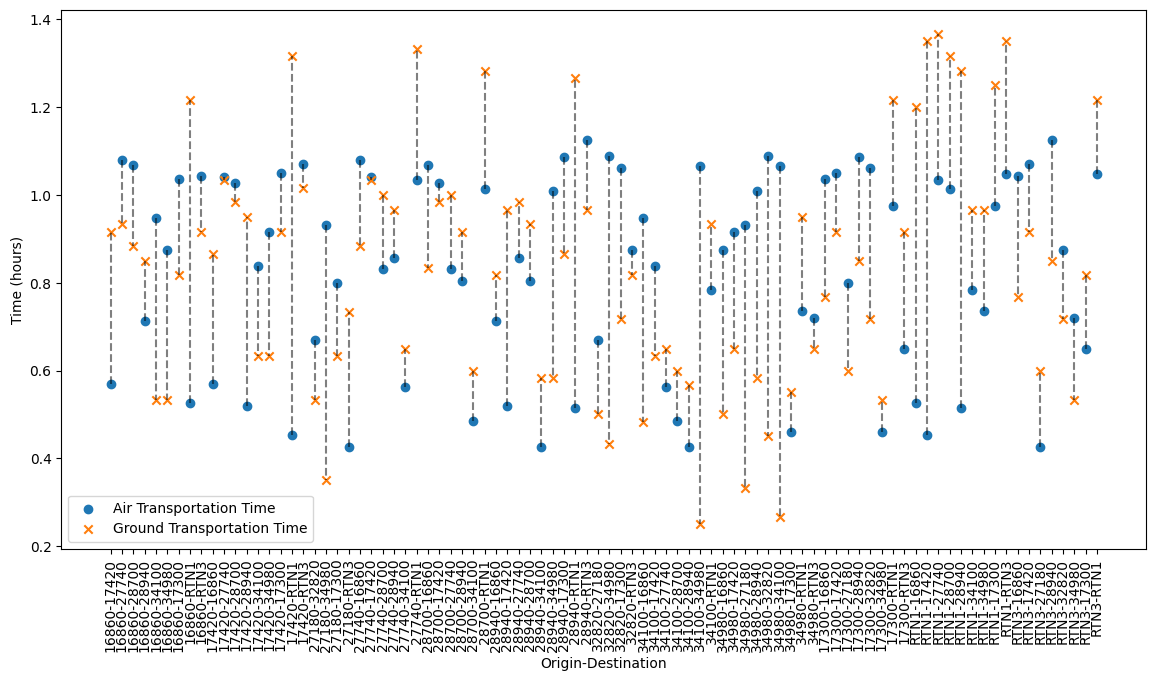}
        \label{fig:Non-RAMTripTime}
    }
    \caption{Air and Ground Transportation Time}
    \label{fig:TripTime}
\end{figure}

\begin{figure}[hbt!]
    \centering
    \subfloat[RAM Trip]{
        \includegraphics[width=0.48\textwidth]{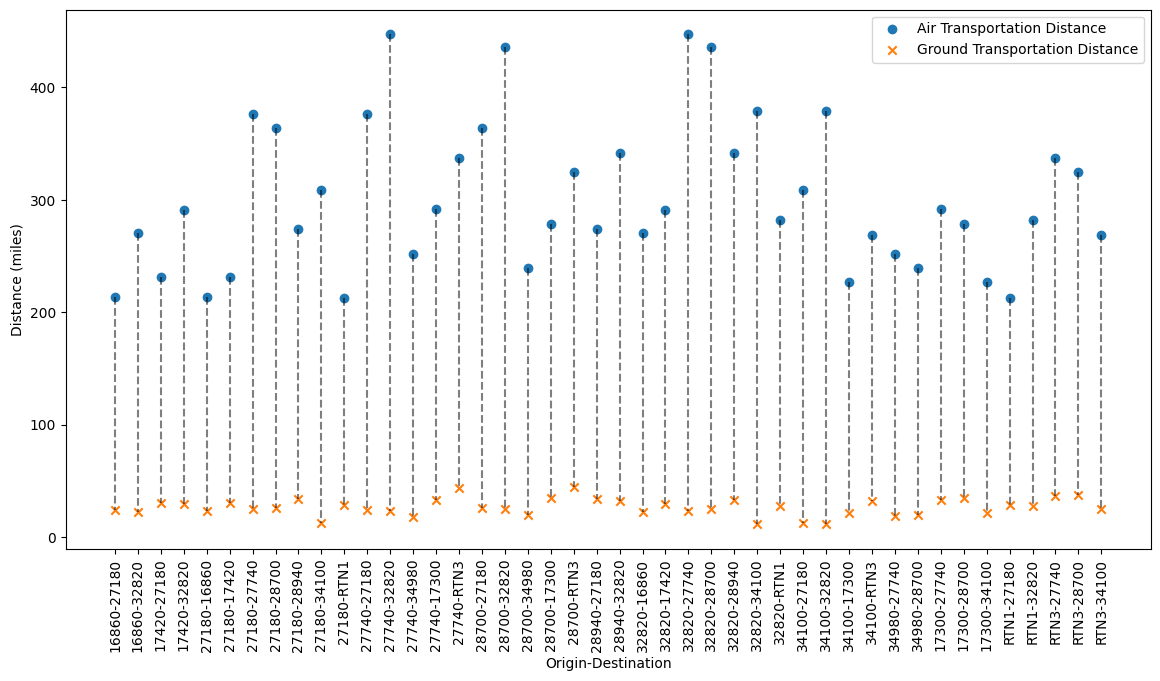}
        \label{fig:RAMTripDistance}
    }
    \hfill
    \subfloat[Non-RAM Trip]{
        \includegraphics[width=0.48\textwidth]{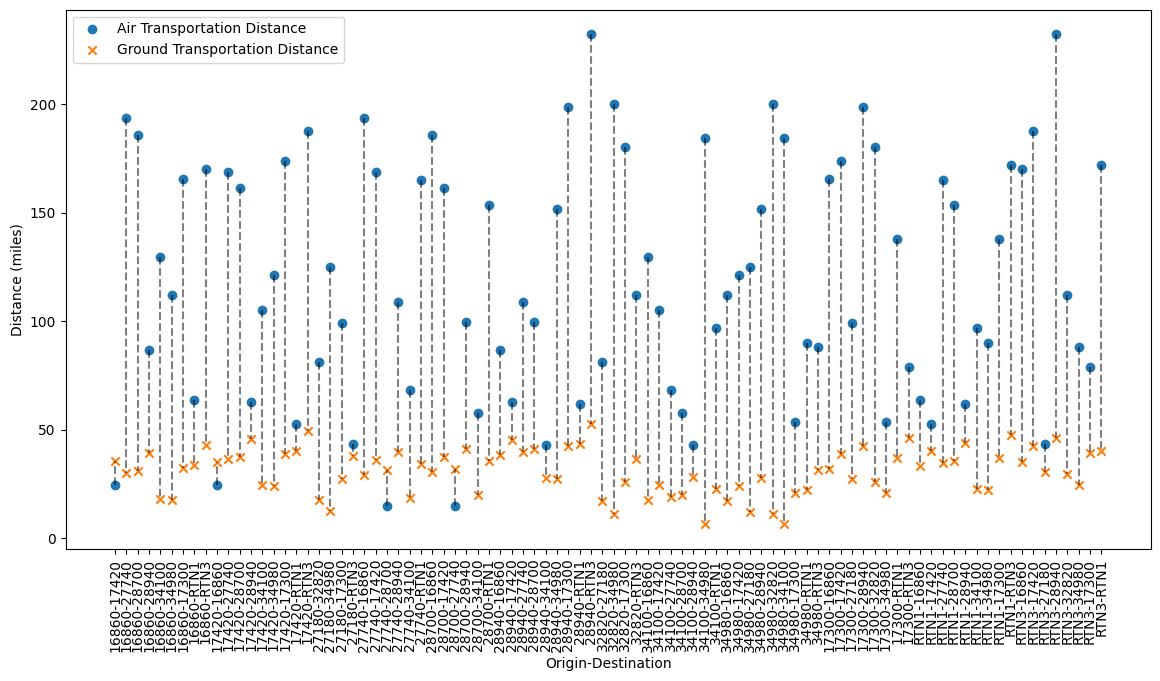}
        \label{fig:Non-RAMTripDistance}
    }
    \caption{Air and Ground Transportation Distance}
    \label{fig:TripDistance}
\end{figure}

As shown in \autoref{fig:TripTime}, a clear distinction is observed between the air and ground transportation times for RAM and non-RAM trips. The Figure \autoref{fig:RAMTripTime} illustrates that for RAM trips, the air transportation time is consistently higher than the ground transportation time, with most air trips exceeding 1 hour. This indicates a trade-off where RAM users prioritize the benefits of air travel despite longer durations. In contrast, Figure  \autoref{fig:Non-RAMTripTime} demonstrates the opposite trend for non-RAM trips, where ground transportation time is typically shorter than air transportation time, reinforcing the preference for ground modes when time efficiency is a critical factor. Similarly, \autoref{fig:TripDistance} highlights the disparity between air and ground transportation distances for RAM and non-RAM trips. In Figure \autoref{fig:RAMTripDistance}, representing RAM trips, the air transportation distance is consistently higher than ground transportation, exceeding 300 miles. This emphasizes the suitability of RAM for longer-distance travel. In contrast, Figure \autoref{fig:Non-RAMTripDistance} illustrates that while air transportation distance is also higher for non-RAM trips, it remains below 200 miles in the majority of instances. This suggests a preference for ground modes in shorter-distance scenarios where the advantages of air travel are less pronounced.

\begin{figure}[hbt!]
\centering
\includegraphics[width=0.5\textwidth]{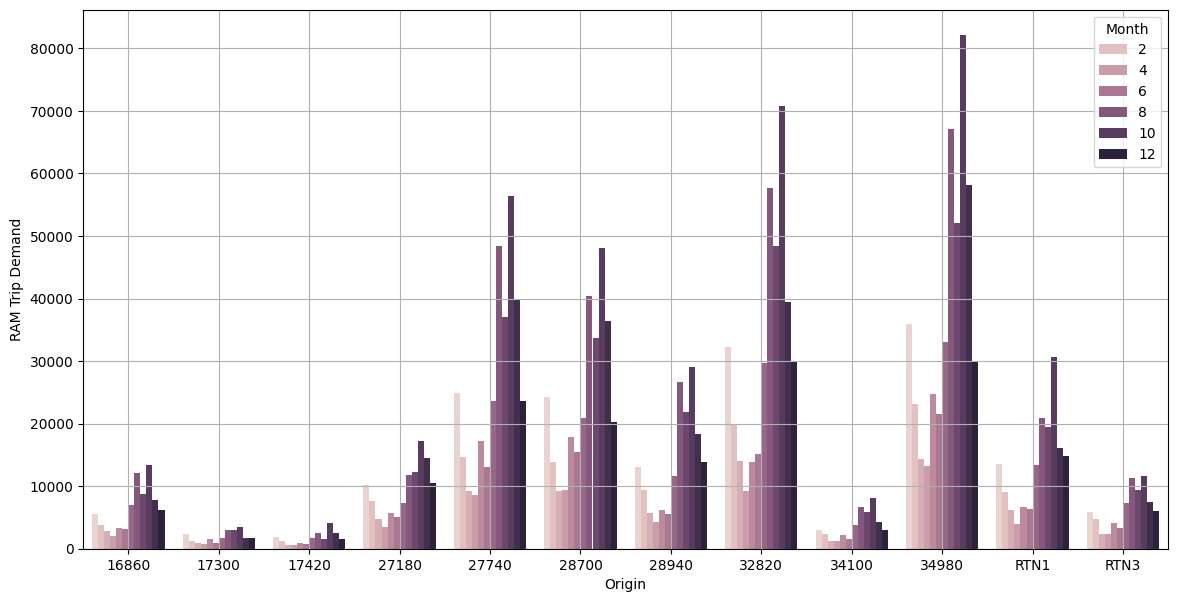}
\caption{Monthly RAM Trip Demand for Each Origin}
\label{fig:monthlyTripDemandOrigin}
\end{figure}

The \autoref{fig:monthlyTripDemandOrigin} highlights that more densely populated MSAs exhibit higher RAM trip demand, as seen at origin points like 27740, 32820, and 34980. These high-demand MSAs consistently show elevated trip counts, particularly during peak months. The demand pattern also varies seasonally, with months later in the year (darker bars representing October, November, and December) typically reflecting increased demand compared to earlier months. This seasonal rise may be attributed to factors such as holiday travel, tourism, and seasonal events.

\begin{figure}[hbt!]
\centering
\includegraphics[height=0.9\textheight]{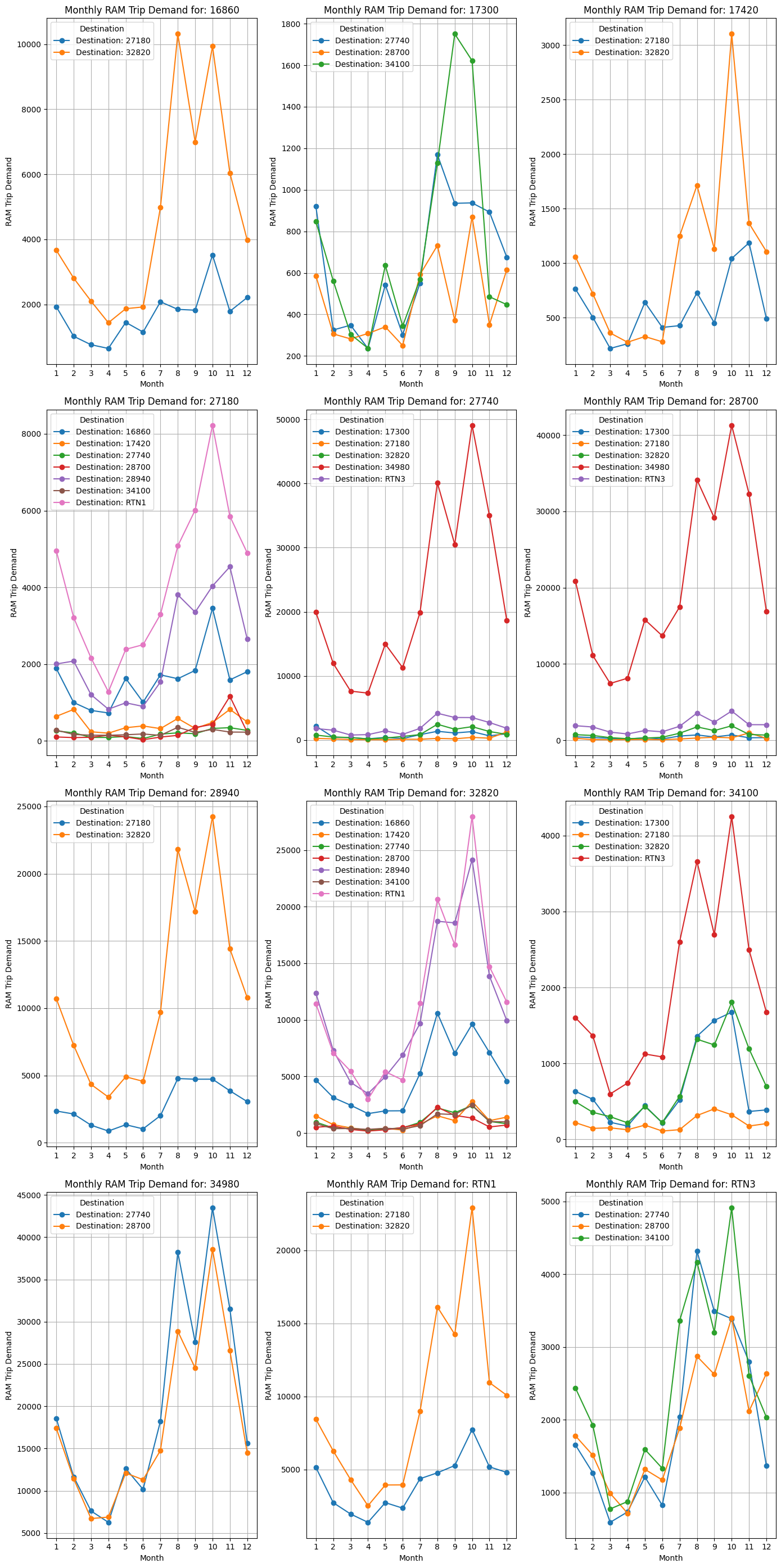}
\caption{Monthly RAM Trip Demand for Each OD pairs}
\label{fig:monthlyTripDemandOD}
\end{figure}

The \autoref{fig:monthlyTripDemandOD} provide a depiction of monthly RAM trip demand across various MSAs in Tennessee. MSAs like 27740 and 28700 demonstrate substantial seasonal spikes in trip demand, particularly toward destinations 34980. This indicates the presence of specific factors, such as tourism, economic activity, or seasonal events, driving heightened mobility. In contrast, other MSAs like 16860 and 17420 exhibit relatively consistent but low demand throughout the year, which could signal limited regional activity or inadequate connectivity to key destinations. The variability in demand patterns highlights the need for a nuanced approach in developing RAM strategies, as blanket solutions may fail to address the unique characteristics of each MSA. Fostering connectivity for underutilized MSAs, improving infrastructure, and diversifying economic drivers could help balance demand across the state and enhance the efficiency of RAM systems in Tennessee.

\section{Conclusion}
This research highlights the potential of RAM as a viable transportation alternative by developing comprehensive cost, time, and risk models. By integrating these into a GTC framework, the study identifies critical factors influencing RAM adoption over ground transportation. For a trip between MSAs in Tennessee, RAM shows promise for trips exceeding 300 miles and scenarios where air travel dominates the GTC, accounting for more than 80\% of the cost. Additionally, OD pairs characterized by GTC value higher than 300 and air travel time of more than 1 hour demonstrate strong potential for shifting to RAM. Seasonal variations in demand highlight the dynamic nature of travel patterns, suggesting the need for adaptive strategies in RAM deployment. Effective planning should prioritize underutilized MSAs to optimize infrastructure investments and ensure balanced resource distribution, paving the way for a sustainable and efficient RAM network.



\section*{Acknowledgments}
This material is based upon work supported by the NASA Aeronautics Research Mission Directorate (ARMD) University Leadership Initiative (ULI) under cooperative agreement number 80NSSC23M0059.

\bibliography{main}

\begin{thebibliography}{23}
\newcommand{\enquote}[1]{``#1''}
\providecommand{\natexlab}[1]{#1}
\providecommand{\url}[1]{\texttt{#1}}
\providecommand{\urlprefix}{URL }
\expandafter\ifx\csname urlstyle\endcsname\relax
  \providecommand{\doi}[1]{\discretionary{}{}{}https://doi.org/#1}\else
  \providecommand{\doi}[1]{\discretionary{}{}{}\urlstyle{rm}\url{https://doi.org/#1}}\fi

\bibitem[{INRIX(2024)}]{INRIX2024}
INRIX, \enquote{INRIX Global Traffic Scorecard,} \url{https://inrix.com/scorecard/}, 2024.
\newblock Accessed: 2024-05-15.

\bibitem[{Johnson and Silva(2022)}]{johnson2022nasa}
Johnson, W., and Silva, C., \enquote{NASA concept vehicles and the engineering of advanced air mobility aircraft,} \emph{The Aeronautical Journal}, Vol. 126, No. 1295, 2022, pp. 59--91.

\bibitem[{Antcliff et~al.(2021)Antcliff, Borer, Sartorius, Saleh, Rose, Gariel, Oldham, Courtin, Bradley, Roy et~al.}]{antcliff2021regional}
Antcliff, K., Borer, N., Sartorius, S., Saleh, P., Rose, R., Gariel, M., Oldham, J., Courtin, C., Bradley, M., Roy, S., et~al., \enquote{Regional air mobility: Leveraging our national investments to energize the American travel experience,} 2021.

\bibitem[{Brink et~al.(2023)Brink, Brown, Carter, Esqué, Meigs, and Riedel}]{mckinseyShorthaulFlying}
Brink, L., Brown, R., Carter, S., Esqué, A., Meigs, B., and Riedel, R., \enquote{{S}hort-haul flying redefined: {T}he promise of regional air mobility --- mckinsey.com,} \url{https://www.mckinsey.com/industries/aerospace-and-defense/our-insights/short-haul-flying-redefined-the-promise-of-regional-air-mobility}, 2023.
\newblock [Accessed 10-06-2024].

\bibitem[{Rajendran and Zack(2019)}]{rajendran2019insights}
Rajendran, S., and Zack, J., \enquote{Insights on strategic air taxi network infrastructure locations using an iterative constrained clustering approach,} \emph{Transportation Research Part E: Logistics and Transportation Review}, Vol. 128, 2019, pp. 470--505.

\bibitem[{Rajendran et~al.(2021)Rajendran, Srinivas, and Grimshaw}]{rajendran2021predicting}
Rajendran, S., Srinivas, S., and Grimshaw, T., \enquote{Predicting demand for air taxi urban aviation services using machine learning algorithms,} \emph{Journal of Air Transport Management}, Vol.~92, 2021, p. 102043.

\bibitem[{Ahmed et~al.(2024)Ahmed, Memon, Rajab, Alshahrani, Abdalla, Rajab, Houe, and Shaikh}]{ahmed2024demand}
Ahmed, F., Memon, M.~A., Rajab, K., Alshahrani, H., Abdalla, M.~E., Rajab, A., Houe, R., and Shaikh, A., \enquote{Demand prediction for urban air mobility using deep learning,} \emph{PeerJ Computer Science}, Vol.~10, 2024, p. e1946.

\bibitem[{Justin et~al.(2021)Justin, Payan, and Mavris}]{justin2021demand}
Justin, C.~Y., Payan, A.~P., and Mavris, D., \enquote{Demand modeling and operations optimization for advanced regional air mobility,} \emph{AIAA Aviation 2021 Forum}, 2021, p. 3179.

\bibitem[{Rimjha et~al.(2021)Rimjha, Hotle, Trani, and Hinze}]{rimjha2021commuter}
Rimjha, M., Hotle, S., Trani, A., and Hinze, N., \enquote{Commuter demand estimation and feasibility assessment for Urban Air Mobility in Northern California,} \emph{Transportation Research Part A: Policy and Practice}, Vol. 148, 2021, pp. 506--524.

\bibitem[{Rimjha et~al.(2022)Rimjha, Hotle, Trani, and Hinze}]{rimjha2022impact}
Rimjha, M., Hotle, S., Trani, A., and Hinze, N., \enquote{Impact of Airspace Restrictions on Urban Air Mobility Commuter Demand Potential,} \emph{2022 Integrated Communication, Navigation and Surveillance Conference (ICNS)}, IEEE, 2022, pp. 1--12.

\bibitem[{Qu et~al.(2024)Qu, Huang, Li, and Liao}]{qu2024demand}
Qu, W., Huang, J., Li, C., and Liao, X., \enquote{A demand forecasting model for urban air mobility in Chengdu, China,} \emph{Green Energy and Intelligent Transportation}, Vol.~3, No.~3, 2024, p. 100173.

\bibitem[{Roy et~al.(2021)Roy, Maheshwari, Crossley, and DeLaurentis}]{roy2021future}
Roy, S., Maheshwari, A., Crossley, W.~A., and DeLaurentis, D.~A., \enquote{Future regional air mobility analysis using conventional, electric, and autonomous vehicles,} \emph{Journal of Air Transportation}, Vol.~29, No.~3, 2021, pp. 113--126.

\bibitem[{Bridgelall(2023)}]{bridgelall2023forecasting}
Bridgelall, R., \enquote{Forecasting market opportunities for urban and regional air mobility,} \emph{Technological Forecasting and Social Change}, Vol. 196, 2023, p. 122835.

\bibitem[{Villa et~al.(2023)Villa, Morejon~Ramirez, and Moore}]{villa2023business}
Villa, I., Morejon~Ramirez, L.~A., and Moore, M., \enquote{The Business Case for Regional Air Mobility at Scale,} \emph{AIAA AVIATION 2023 Forum}, 2023, p. 3295.

\bibitem[{Justin et~al.(2022)Justin, Payan, and Mavris}]{justin2022integrated}
Justin, C.~Y., Payan, A.~P., and Mavris, D.~N., \enquote{Integrated fleet assignment and scheduling for environmentally friendly electrified regional air mobility,} \emph{Transportation Research Part C: Emerging Technologies}, Vol. 138, 2022, p. 103567.

\bibitem[{Acharya et~al.(2024)Acharya, Lad, Sun, and Song}]{acharya2024demandmodelingadvancedair}
Acharya, K., Lad, M., Sun, L., and Song, H., \enquote{Demand Modeling for Advanced Air Mobility,} , 2024.
\newblock \urlprefix\url{https://arxiv.org/abs/2412.06807}.

\bibitem[{{Federal Highway Administration}(2021-22)}]{FHWA2022}
{Federal Highway Administration}, \enquote{2022 NextGen NHTS National Passenger OD Data,} U.S. Department of Transportation, Washington, DC. Available online: \url{https://nhts.ornl.gov/od/}, 2021-22.

\bibitem[{Helmer(2008)}]{Helmer2008MicropolitanSA}
Helmer, G., \enquote{Micropolitan Statistical Areas: A Few Highlights,} \emph{Monthly Labor Review}, Vol. 131, 2008, p.~40.
\newblock \urlprefix\url{https://api.semanticscholar.org/CorpusID:158508202}.

\bibitem[{{U.S. Department of Transportation}(2024)}]{USDOT2024}
{U.S. Department of Transportation}, \enquote{Revised Departmental Guidance on Valuation of a Statistical Life in Economic Analysis,} \url{https://www.transportation.gov/office-policy/transportation-policy/revised-departmental-guidance-on-valuation-of-a-statistical-life-in-economic-analysis}, 2024.
\newblock Accessed: 2024-06-10.

\bibitem[{Soe and Thein(2020)}]{soe2020haversine}
Soe, N.~C., and Thein, T. L.~L., \enquote{Haversine formula and RPA algorithm for navigation system,} \emph{International Journal of Data Science and Analysis}, Vol.~6, No.~1, 2020, p.~32.

\bibitem[{De~Blaeij et~al.(2003)De~Blaeij, Florax, Rietveld, and Verhoef}]{de2003value}
De~Blaeij, A., Florax, R.~J., Rietveld, P., and Verhoef, E., \enquote{The value of statistical life in road safety: a meta-analysis,} \emph{Accident Analysis \& Prevention}, Vol.~35, No.~6, 2003, pp. 973--986.

\bibitem[{Tanner(1981)}]{tanner1981expenditure}
Tanner, J.~C., \enquote{Expenditure of time and money on travel,} \emph{Transportation Research Part A: General}, Vol.~15, No.~1, 1981, pp. 25--38.

\bibitem[{Heiss(2016)}]{heiss2016discrete}
Heiss, F., \enquote{Discrete Choice Methods with Simulation,} \emph{Econometric Reviews}, Vol.~35, No.~4, 2016, pp. 688--692.

\end{thebibliography}

\end{document}